\documentclass[useAMS,usenatbib]{mn2e}
\usepackage{graphicx}
\usepackage{epstopdf}
\usepackage{natbib}
\usepackage{aas_macros}
\usepackage{color}
\defcitealias{Hoi2013}{Horvath et al.}

\title{{\bf A giant ring-like structure at $0.78<z<0.86$
displayed by GRBs}}
\author[L. G. Bal\'azs et al.]{L. G. Bal\'azs$^{1,2}$\thanks{E-mail,
balazs@konkoly.hu}, Z. Bagoly$^{2,3}$, J. E. Hakkila$^{4}$, I.
Horv\'ath$^{3}$, J. K\'obori$^{2}$, \newauthor I. R\'acz$^{1}$, L. V. T\'oth$^{2}$ \\
\\
$^{1}$MTA CSFK Konkoly Observatory, Konkoly-Thege M. \'ut 13-17, Budapest, 1121, Hungary\\
$^{2}$E\"otv\"os University, P\'azm\'any P\'eter s\'et\'any 1/A,
Budapest,1117, Hungary\\
$^{3}$National University of Public Service, 1083, Budapest,
Hungary \\
$^{4}$Department of Physics and Astronomy, The College
of Charleston, Charleston, SC 29424-0001, USA }

\begin{document}
\date{}
\pagerange{\pageref{firstpage}--\pageref{lastpage}} \pubyear{}
\maketitle
\label{firstpage}

\begin{abstract}
According to the cosmological principle, Universal large-scale
structure is homogeneous and isotropic. The observable Universe,
however, shows complex structures even on very large scales. The
recent discoveries of structures significantly exceeding the
transition scale of 370 Mpc pose a challenge to the cosmological
principle.

We report here the discovery of the largest regular formation in
the observable Universe; a ring with a diameter of 1720 Mpc,
displayed by 9 gamma ray bursts (GRBs), exceeding by a factor of
five the transition scale to the homogeneous and isotropic
distribution. The ring has a major diameter of $43^o$ and a minor
diameter of $30^o$ at a distance of 2770 Mpc in the $0.78 < z <
0.86$ redshift range, with a probability of $2\times 10^{-6}$ of
being the result of a random fluctuation in the GRB count rate.

Evidence suggests that this feature is the projection of a shell
onto the plane of the sky. Voids and string-like formations are
common outcomes of large-scale structure. However, these
structures have maximum sizes of 150 Mpc, which are an order of
magnitude smaller than the observed GRB ring diameter. Evidence in
support of the shell interpretation requires that temporal
information of the transient GRBs be included in the analysis.

This ring-shaped feature is large enough to contradict the
cosmological principle.
The physical mechanism responsible for causing it
is unknown.
\end{abstract}

\begin{keywords}
Large-scale structure of Universe, cosmology: observations,
gamma-ray burst: general
\end{keywords}

\section{Introduction}
Quasars are well-suited for mapping out the large-scale
distribution of matter in the Universe, due to their very high
luminosities and preferentially large redshifts. Quasars are
associated by groups and poor clusters  of galaxies
\citep{Hei2005,Lie2009} and can be observed even when the
underlying galaxies are faint and difficult to detect. When
quasars cluster, they identify considerable amounts of underlying
matter, such that quasar clusters have been used to detect matter
clustered on very large scales. Some of this matter is clustered
on scales equal to or exceeding that of the Sloan Great Wall
\citep{Got2005}.

A number of large quasar groups (LQG) have been identified in
recent years; each one mapping out large amounts of much fainter
matter. After \cite{Web1982} found a group of four quasars at $z =
0.37$ with a size of about 100 Mpc, having a low probability of
being a chance alignment, \cite{Kom1994} identified strong
clustering in the quasar distribution at scales less than $20\,
h^{-1}$ Mpc, and defined LQGs using a well-known cluster analysis
technique. Subsequently, \cite{Kom1996} identified additional
LQGs, and \cite{Kom1998} reported a new finding of eleven LQGs
based on systematic cluster analysis. The sizes of these clusters
ranged from 70 to 160 $h^{-1}$ Mpc. \cite{New1998b,New1998a} later
discovered a $~150 \, h^{-1}$ Mpc group of 13 quasars at median
redshift $z \cong 1.51$.
 \cite{Wil2002} mapped 18 quasars spanning $\approx 5^\circ \times
2.5^{\circ}$ on the sky, with a quasar spatial overdensity $6-10$
times greater than the mean. \cite{Hab2009} investigated two
sheet-like structures of galaxies at $z = 0.8$ and $1.3$ spanning
$150 \, h^{-1}$ comoving Mpc embedded in LQGs extending over at
least $200 \, h^{-1}$ Mpc. \cite{Hai2004} reported the finding of
two large-scale structures of galaxies in a $40 \times 35$
arcmin$^2$ field embedded in a $25^{\circ 2}$ area containing two
100 Mpc-scale structures of quasars. As the identified scales of
quasar clusters became larger, \cite{clo2012} found two relatively
close LQGs at $z \sim 1.2$. The characteristic sizes of these two
LQGs, $\approx 350-400$ Mpc, and appear to be only marginally
consistent with the scale of homogeneity in the concordance
cosmology. Recently, \cite{clo2013} uncovered the Huge-LQG with a
long dimension of $\approx 1240$ Mpc ($1240 \times 640 \times 370$
Mpc). Until recently, this was the largest known structure in the
Universe. Using a friend of friend (FoF) algorithm \cite{Ein2014}
found that the linking length $l = 70 \, h^{-1} Mpc$ three systems
from their QSR catalogue coincide with the LQGs
 from \cite{clo2012,clo2013}.

 \cite{Hoi2013,Hoi2014} announced the discovery of
a larger Universal structure than the Huge-LQG by analyzing the
spatial distribution of gamma-ray bursts (GRBs). The 3000 Mpc size
of this structure exceeds the size of the Sloan Great Wall  by a
factor of about six; This is currently the largest known universal
structure.

Unlike quasars, GRBs are short-lived cosmic transients spanning
milliseconds to hundreds of seconds \citep[see the review paper
by][]{Mes2006}. Due to their immense luminosities, GRBs can be
observed at very large cosmological distances. Their hosts are
typically metal poor galaxies of intermediate mass
\citep{Sav2009,Cas2010,Lev2010}, rather than the massive galaxies
in which quasars are generally found. Both quasars and GRBs should
map the underlying distribution of universal matter, although the
details of their spatial distributions are not necessarily
identical. Although the number of known GRBs for which distances
have been accurately measured is significantly fewer than the
number of known quasars, the surveying techniques used to identify
these objects are more homogeneous and better-suited for studying
structures of large angular size than are quasars.

The existence of an object with a size of several gigaparsecs
introduces questions concerning the homogenous and isotropic
nature of cosmological models. The great importance of this question
necessitates further independent study into the use of GRBs for
mapping large-scale universal structures.
Our analysis considers the space distribution of
a GRB sample having known redshifts for the
presence of large-scale anisotropies. The sample we
use for this study is available at {\tt
http://www.astro.caltech.edu/grbox/grbox.php}.
As of October 2013, the redshifts of 361 GRBs have been determined.

\section{Distribution of GRBs in $\{r,\theta, \varphi \}$ space}
According to the cosmological principle ($CP$), Universal
large-scale structure is homogeneous and isotropic
\citep{Ell1975}. The WMAP and Planck experiments have revealed
that the Big Bang had these properties in its early expansion
phase. The observable Universe, however, shows complex structures
even on very large scales. The problem is to find a
limiting scale at which the $CP$ is valid.

A number of well-known studies have attempted to find the largest
scale on which $CP$ is valid. According to \cite{Ein1993} the
available data suggested values $r(t) = 130 \pm 10 \, h^{-1}$.
\cite{Yah2005} reported on the first result from the clustering
analysis of SDSS quasars: the bump in the power spectrum due to
the baryon density was not clearly visible, and they concluded
that the galaxy distribution was homogeneous on scales larger than
$60-70 \, h^{-1}$ Mpc. \cite{Teg2006} using luminous red galaxies
(LRGs) in the Sloan Digital Sky Survey (SDSS) improved the
evidence for spatial flatness ($\Omega_{tot}=1$). \cite{Lii2012}
have constructed a set of supercluster catalogues for the galaxies
from the SDSS survey main and LRG flux-limited samples.
\cite{Bag2008} showed that in the concordance model, the fractal
dimension makes a rapid transition to values close to 3 at scales
between 40 and 100 Mpc. \cite{Sar2009} found the galaxy
distribution to be homogeneous at length-scales greater than $70
\, h^{-1}$ Mpc, and \cite{Yad2010} estimated the upper limit to
the scale of homogeneity as being close to $260 \, h^{-1}$ Mpc for
the $\Lambda CDM$ model. \cite{Soh2012} studied the Ultra Deep
Catalogue of Galaxy Structures; the cluster catalogue contains
1780 structures covering the redshift range $0.2 < z < 3.0$,
spanning three orders of magnitude in luminosity  $(10^8 < L_4 < 5
\times 10^{11} L_\odot)$ and richness from eight to hundreds of
galaxies. These results supported the validity of $CP$.

%\cite{Yad2010} estimated the upper limit to the scale of
%homogeneity as being close to $260 \, h^{-1}Mpc$ for the $\Lambda CDM$
%model.
%\cite{Soh2012} studied the Ultra Deep Catalogue of Galaxy
%Structures; the cluster catalogue contains 1780 structures covering the
%redshift range $0.2 < z < 3.0$, spanning three orders of magnitude
%in luminosity  $(10^8 < L_4 < 5 \times 10^{11} L_\odot)$ and
%richness from eight to hundreds of galaxies. \cite{Sar2009} found
%the galaxy distribution to be homogeneous at length-scales
%greater than $70h^{-1}$ Mpc. \cite{Bag2008} showed that in the
%concordance model, the fractal dimension makes a rapid transition
%to values close to 3 at scales between 40 and 100 Mpc.
%\cite{Yad2005} concluded that the galaxy distribution was
%homogeneous on scales larger than $60-70h^{-1}$ Mpc. According to
%\cite{Ein1993} the available data suggested values $r(t) = 130 \pm
%10 h^{-1}$. \cite{Teg2006} using luminous red galaxies (LRGs) in
%the Sloan Digital Sky Survey (SDSS) improves the evidence for
%spatial flatness ($\Omega_{tot}=1$). \cite{Eis2005} find a
%well-detected peak in the correlation function at $100 h^{-1}$
%Mpc the imprint of the recombination-epoch acoustic
%oscillations. \cite{Yah2005} reported on the first result from the
%clustering analysis of SDSS quasars. The bump in the power
%spectrum due to the baryon density was not clearly visible.
%These results supported the validity of $CP$.

Assuming the validity of $CP$, a homogeneous isotropic model and
a standard $\Lambda C D M$ cosmology ($\Omega_\Lambda=0.7,
\Omega_M=0.3, h=0.7$, representing an Euclidean
space with $\Omega_\Lambda+\Omega_M=0.7+0.3=1$) the line
element in the $4D$ space-time is given by

\begin{equation}
dl^2=R(t)^2(dr^2+r^2d\vartheta^2+r^2sin\vartheta^2
d\varphi^2)-c^2dt^2
\end{equation}

\noindent The variables in the equation have their conventional
meaning.  The change of the spatial part of $dl^2$ line element in
course of time is given by the $R(t)$ scale factor so the spatial
distance in the brackets i.e.

\begin{equation}
ds^2=dr^2+r^2d\vartheta^2+r^2sin\vartheta^2 d\varphi^2
\end{equation}

\noindent is independent of the time. Any event in the $4D$
space-time has a footprint in the $\{r, \theta, \varphi \}$ space
where $r$ can be computed from the observed redshift $z$ and the
angular coordinates are given by the observations. In
astronomy the angular coordinates are usually concretized in
equatorial or Galactic systems. The $r$ distance is measured by
the comoving distance defined in the Euclidean case by

\begin{equation}
r(z)=\frac{c}{H_0} \int^z_0
\frac{dz'}{\sqrt{\Omega_M(1+z'^3)+\Omega_\Lambda}}
\end{equation}

\noindent where $c$ is the speed of light and $H_0$ is the Hubble
constant.

The distribution of GRBs in the
$\{r,\theta, \varphi \}$ coordinate system can be constructed by assuming some
universal formation history, along with spatial homogeneity and isotropy. This
theoretical distribution, however, cannot be observed directly
because the observations are biased by selection effects.
There are several factors influencing the probability that a GRB
is detected. The limit of the instrumental sensitivity is such a
factor, as GRBs below this threshold cannot be detected. The
probability of detecting a GRB depends also on the observational
strategy of the satellite.

The GRBs in our sample having known redshift were detected by different
satellites using different observational strategies.
The method of observation results in
different detection probabilities (known as the {\em exposure function}) since each
satellite spends different time durations observing various parts of
the sky. In principle this exposure function can be reconstructed
from the log of observations made by each satellite. The GRB redshift
can be obtained either from spectral observation of the GRB
afterglow or of the host galaxy if it can be localized.
The exposure is a function of GRB brightness and
not of GRB redshift.

Redshift measurements have their own biases.
These depend on the
optical brightness of the afterglow or the host galaxy, depending on which
is observed. The most significant factor influencing
the possibility of observation is extinction caused by the Galactic
foreground. This bias seriously influences the probability of
redshift measurement. However, galactic extinction
has a known distribution and can be estimated at cosmological
distances for
any angular position on the sky.

Most GRB afterglows and host galaxies are optically faint so
one needs large-aperture telescopes in order to measure GRB redshifts.
Northern and Southern hemisphere telescopes used to make these
measurements are located at
mid-latitudes; consequently the chance of getting the necessary
observing time is higher in the winter than in the summer since
the night is much longer during this season. That part of
the sky which is accessible in the winter season has a higher
probability for determining a GRB redshift. This bias is equalized
when observations are made over many observing seasons.
However, since host galaxy measurements can be made at
any time after the GRB is observed, this effect is less important
when measuring the redshift from the host.

The net effect of all these factors
can be expressed by the following formula:

\begin{equation}
f_{obs}(\vartheta,\varphi,z)=T_{tel}(\vartheta,\varphi)E_{xt}(\vartheta,
\varphi)E_{xp}(\vartheta,\varphi)f_{int}(z)
\end{equation}

\noindent The $T_{tel}$ telescope time and $E_{xp}$  exposure
function factors do not depend on the redshift. As to the $E_{xt}$
extinction, however, one may have some concerns. If the observable
optical brightness depends on the distance, which seems to be a
quite reasonable assumption, then the measured $z$ values close to
the Galactic equator could be systematically lower. Our sample of
361 GRBs, however, does not show this effect. One may assume,
therefore, that the observed distribution of GRBs in the
$\{r,\theta, \varphi \}$ space can be written in the form of

\begin{equation}\label{GF}
f_{obs}(\vartheta,\varphi,z)=g(\vartheta,\varphi)f_{int}(z)
\end{equation}

\noindent In deriving this equality it is assumed that
$f_{int}$ does not depend on the $\vartheta,\varphi$
angular coordinates, due to isotropy.

\section{Testing the angular isotropy of the GRB distribution}
A number of approaches have been developed for studying the
non-random departure
from the homogeneous isotropic distribution of matter in the
universe. Each method
is sensitive to different forms of the departure of the cosmic
matter density from the homogeneous isotropic case. These
tests have been previously applied to galaxy and quasar
distributions.
\cite{Clo1987} described a
three-dimensional clustering analysis of 1100 'high-probability'
quasar candidates occupying the assigned-redshift band of 1.8 - 2.4.
\cite{Ick1991} presented a geometrical model, making use the
Voronoi foam, for the asymptotic distribution of the cosmic mass
on 10-200 Mpc scales.
\cite{Gra1995} applied a graph theoretical method, using the
minimal spanning tree (MST), to find candidates for quasar
superstructures in quasar surveys.
\cite{Dor2004} used the MST technique to extract sets of
filaments, wall-like structures,
galaxy groups, and rich clusters.
\cite{Pla2011} studied the linear Delaunay Tessellation Field
Estimator (DTFE), its higher order equivalent Natural Neighbour
Field Estimator (NNFE), and a version of the Kriging interpolation.
The DTFE, NNFE and Kriging approaches had largely similar density and
topology error behaviours. \cite{Zha2010} introduced a method
for constructing galaxy density fields based on a Delaunay
tessellation field estimation (DTFE). Recently, \cite{Kit2013,kit2014}
presented the KIGEN-approach  which allows for the first time for
self-consistent phase-space reconstructions
from galaxy redshift data.

Even if $CP$ is correct, the distribution of GRBs in the comoving
frame will likely be heterogeneous due to the cosmic history of
structure formation. However, the isotropy of the angular
distribution can remain unchanged. Based on the data obtained by
the CGRO BATSE experiment, \cite{Bri1993} concluded that angular
distribution of GRBs is isotropic on a large scale. In contrast,
later studies performed on larger samples suggest that this is
true only for bursts of long duration ($>10 \, sec$) rather than
subclasses of GRBs having different durations and presumably
different progenitors
\citep{Bal1998,Bal1999,Mes2000,Lit2001,Tan2005,Vav2008}.

\subsection{Testing the isotropy based on conditional probabilities}
The validity of Equation (\ref{GF}) can be tested by some
suitably-chosen statistical procedure. The equation indicates that the
distribution of the angular coordinates is independent of the $z$
redshift. By definition this independence also means that the
$g(\vartheta,\varphi|z)$ conditional probability density of the
angular distribution, assuming that $z$ is given, does not depend
on $z$, i.e. the joint probability density of the angular and
redshift variables can be written in the form

\begin{equation}\label{FGF}
f_{obs}(\vartheta,\varphi,z)=g(\vartheta,\varphi|z)f_{int}(z)=g(\vartheta,\varphi)f_{int}(z)
\end{equation}

\noindent Let us consider an $\{\vartheta_i,\varphi_i,z_i;
i=1,2,\ldots,n \}$ observed sample of GRB positional and redshift data.
The validity of Equation (\ref{FGF}) can be tested on this dataset in
different ways. One approach is to test whether or not the conditional
probability is independent of the condition. This is the approach
used by \citetalias{Hoi2013}, who split the sample into $k$ subsamples
according to the $z$-characteristics and tested whether or not
the subsamples could originate from the same $g(\vartheta,\varphi)$.
This is the null hypothesis to be tested. If the null hypothesis is true,
then the way in which the sample is subdivided by $z$ is not crucial.
A reasonable approach is to select the subsamples so that they have
equal numbers of GRBs; this ensures the same statistical properties of the subsamples.
The number of subsamples $k$ is somewhat arbitrary.

\citetalias{Hoi2013} selected $k$ different radial bins and
performed several tests for sample isotropy on each bin.
The tests singled out the slice at $1.6 \le z <2.1$ as having a
statistically significant sample anisotropy, which was due
to a large cluster of GRBs in the northern Galactic sky.

\subsection{Testing the isotropy of the GRB distribution using joint probability factoring}
It is important to independently test the significance of the
\citetalias{Hoi2013} result using alternate statistical methods.
We choose to regard the GRB distribution
in terms of Equation (\ref{FGF}).
The right side of Equation
(\ref{FGF}) describes a factorization of the
$f_{obs}(\vartheta,\varphi,z)$ joint probability density in terms
of the angular and redshift variables. To test the
validity of this factorization we proceed in the following way:
we consider again the sample of $\{\vartheta_i,\varphi_i,z_i;
i=1,2,\ldots,n \}$ mentioned above. If the factorization
is valid (e.g. if the angular distribution is independent of redshift),
then the sample is invariant under a random resampling of the $z$ variable
while keeping the angular coordinates unchanged.
Using this approach we get a new sample that is statistically equivalent
to the original one, assuming that the joint probability
density factorization is valid.

The sample coming from the $f_{obs}(\vartheta,\varphi,z)$ joint
probability density is three dimensional, so the task at hand is a
comparison of three-dimensional samples. One way in which the
sample can deviate from isotropy is through the presence of one or
more density enhancements and/or decrements. For this type of
density perturbation $f_{int}$ is also dependent on the angular
coordinates and the factorization is no longer valid. In this
case, the resampling of the redshifts may change $f_{obs}$ as
well. Comparing the original observed distribution with the
resampled one can identify the presence of density perturbations
with respect to the isotropic case.

To calculate the Euclidean distances in a simple way we introduce
$x_c,y_c,z_c$ Descartes coordinates in the comoving
$r,\vartheta,\varphi$ frame. Using Galactic coordinates one
obtains

\begin{eqnarray}
x_c&=& r\, cos(b)cos(l) \\
y_c&=& r\, cos(b)sin(l) \\
z_c&=& r\, sin(b)
\end{eqnarray}

\noindent Having a sample of size $N$ the number of objects in a
differential volume $dV$ can be written using Descartes
coordinates in the form $dN=\nu (x,y,z)dV=\nu (x,y,z)dxdydz$,
where $\nu$ denotes the spatial density of the objects. It is
clear from this formula that $\nu(x,y,z) = N f_{int}(x.y.z)$.
Obviously, $\nu (x,y,z)=dN/dV$.

A trivial estimate of $\nu$ is obtained by counting $dN$ in a given
$dV$. Fixing $dV$, the variance of $\nu$ is given by $dN$. An
alternative approach is to keep $dN$ constant and look
for the appropriate $dV$. This approach can be realized by
computing the distance to the $k$-th nearest neighbour in the
sample. The distance is the conventional Euclidean one in our
case. Proceeding in this way, $\nu (x,y,z) = 3(k+1)/(4\pi r_k^3)$
where $r_k$ is the distance to the $k$-th nearest neighbour. In the
following subsection we adopt this approach for the GRB sample.

\subsection{$k$-th Next neighbor statistics for the GRB data}
\label{knear}
In order to find the $k$-th next neighbours we use
the \textcolor{red}{\tt knn.dist(x,k)} procedure in the {\it FNN}
library of the {\bf R} statistical
package\footnote{http://cran.r-project.org}. Given $k$, the
procedure computes the distances of the $x$ sample elements up to
the $k$-th order. Resampling the data $r$ can proceed by using the
\textcolor{red}{\tt sample(x,n)} procedure of {\bf R} where $x$
refers to the dataset being resampled and $n$ is the size of the
new sample (which is identical to the original one in our case).
We have made 10000 resamplings of the comoving distance sample and
computed the densities choosing $k= 1,2,\ldots ,20$ for the
nearest-neighbour distances. The densities obtained from the
resampled data enable us to compute the mean density and its
variance at each of the sample distances.

\begin{figure}
  \centering
  \includegraphics[width=8cm]{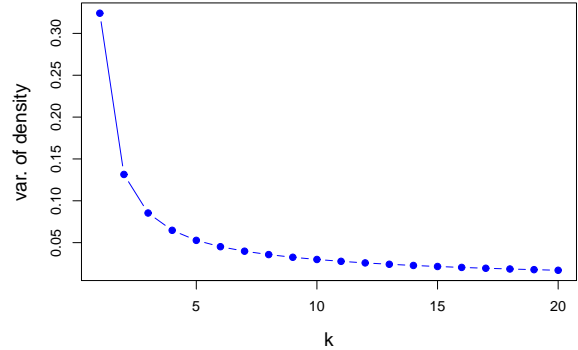}\\
  \caption{Dependence of the density variance on the $k$-th order of the
  nearest neighbour. Note the rapid decrease of variance at lower $k$ and
  the much shallower decrease for larger values.}\label{vark}
\end{figure}

At this point the value of $k$ is quite arbitrary. Selecting a small
value of $k$ might make this procedure sensitive to small scale
disturbances. However, the variance of the estimated density
on this scale is large.
In contrast, the variance of the density is lower for
larger values of $k$ but the density fluctuations of smaller scale
might be smeared out. In order to make a reasonable compromise between
$k$ and the variance of the estimated density, we compute the mean
variance in the function of $k$. The result is displayed in
Figure~\ref{vark}.

\begin{figure}
  \includegraphics[width=8cm]{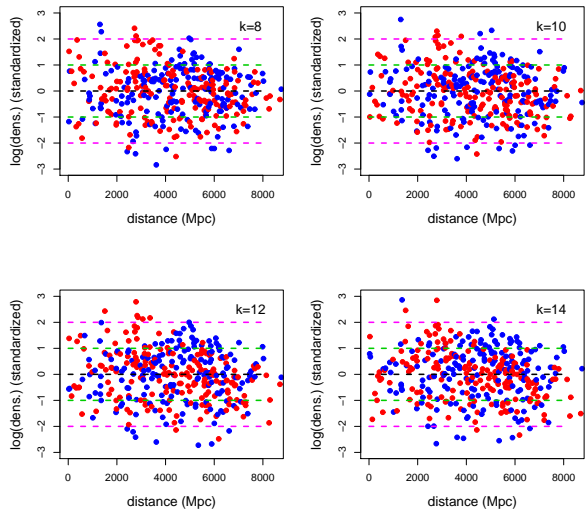}\\
  \caption{Dependence of the standardized logarithmic density on  the
  comoving distance.  The  0~($mean), 1\sigma, 2\sigma$ lines are marked
  with black, green and magenta colors, respectively. The GRBs in the
  Southern Galactic hemisphere and those in the Northern are marked with red
  and blue colors. Note that a group of red points close to the $2\sigma$ line at about
  2800~Mpc may correspond to a real density enhancement of
  GRBs in the Southern Galactic hemisphere. The GRB Great Wall
  discovered by \citetalias{Hoi2013} appears as a group of blue
  points between the $1\sigma$ and $2\sigma$ lines in the
  4000-6000 Mpc distance range.}\label{logd}
\end{figure}

The Figure helps us to estimate the optimal selection of the $k$
value. Up to $k=5$ the variance drops rapidly and starting about
$k=8$ its change is much shallower. We select the values of
$k=8,10,12,14$ and study densities in these four representative cases.
The simulated 10000 runs enable us to calculate the mean and variance
of the density at every GRB location in the sample. Using the variance and
the mean of the density we calculate the standardized (zero mean
and unit variance) values of the density for all points of the
sample. In Figure \ref{logd} we display a scatter plot between
the standardized logarithmic density and the comoving distance for
the cases of $k=8, 10, 12$ and $14$, respectively.

A quick glance at Figure \ref{logd} reveals a group of red
(Southern) dots between the $1\sigma$ and $2\sigma$ lines at about
2800 Mpc. These dots may represent an associated group of GRBs. The Great
GRB Wall discovered by \citetalias{Hoi2013} can be recognized as
an enhancement of blue (Northern) dots between the $1\sigma$ and
$2\sigma$ lines in the 4000-6000 Mpc distance range. These
impressions, however, are somewhat subjective, their validity needs
to be confirmed by some suitable statistical study for a more
formal significance.

The null hypothesis in this case is the assumption that
all spatial density enhancements of GRBs are produced purely by
random fluctuations. Assuming the validity of the null hypothesis
one has to compute the probability of the density enhancement in
question. We perform a series of Kolmogorov-Smirnow (KS) tests and
confirm that the standardized logarithmic densities displayed in Figure
\ref{logd}
 follow a Gaussian distribution. Denoting the logarithmic density
 by $\varrho$,
 %the $i$-th element's probability is $p_i
%\propto exp(-\varrho^2_i/2)$ and the log-likelihood function has
the probability that the $i$-th element's value is a density
enhancement is $p_i
\propto exp(-\varrho^2_i/2)$ and the log-likelihood function has
the form

\begin{equation}\label{lf}
    L= -\frac{1}{2} \sum \limits^n_{i=1} \varrho^2_i+ const.
\end{equation}

 \noindent The summation in Equation (\ref{lf}) yields
a $\chi^2_n$ variable with $n$ degrees of freedom. All density
fluctuations are restricted within certain distance ranges. To get a
likelihood function sensitive to a given density
enhancement and/or deficit we have to sum up successive points in
Equation (\ref{lf}).  Taking and summing up $k$ successive points
within a distance range we get a $\chi^2_k$ variable having $k$ degrees of
freedom. All density values are calculated from a fixed number of
nearest neighbours in the sample. We choose for $k$ the order of the
nearest value from which the density is calculated.

There is some arbitrariness to this procedure. The $k$ orders of
the next neighbours have been selected somewhat by insight.
Nevertheless, the calculated densities correlate strongly. This
property enables us to concentrate the densities into one
variable.

\begin{table}
  \centering
  \begin{tabular}{|c|cccc|}
    \hline
    order  & k=8 & k=10 & k=12 & k=14 \\
    \hline
    k= 8 & 1.000 &  0.914&  0.540&  0.706\\
    k=10 &  0.914&  1.000 &  0.515 &  0.661\\
    k=12 & 0.540 &  0.515&  1.000&  0.743\\
    k=14 & 0.706 & 0.661 &  0.743&  1.000\\
    \hline
  \end{tabular}
  \caption{Correlation between logarithmic densities computed from
  $k^{th}$ orders of nearest neighbours.}\label{co}
\end{table}

The correlation matrix in Table \ref{co} reveals a strong
correlation between nearest neighbor estimation of densities of
the orders
$k=8$, $k=10$, $k=12$ and $k=14$. To get a joint variable
we perform principal component analysis (PCA) on the
standardized logarithmic densities used above. To get the PCs we
use the \textcolor{red}{\tt princomp()} procedure from {\bf R}'s
{\it stats} library.

\begin{figure}
  \includegraphics[width=8cm]{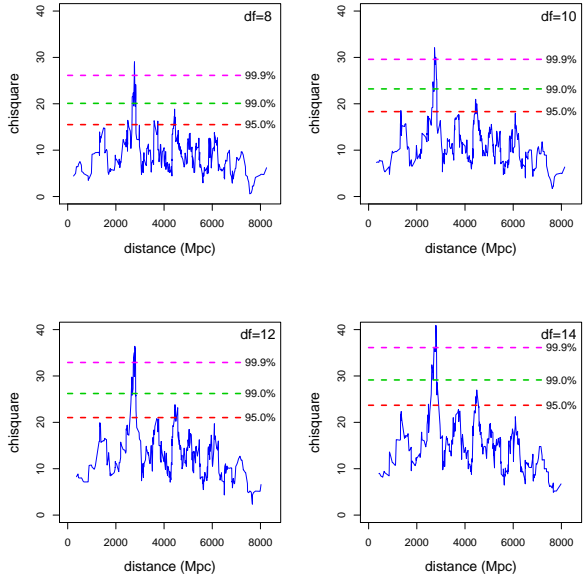}\\
  \caption{The calculated $\chi^2_k$ values using the first PC. Their degree of freedom is given
  in the right top corner of the corresponding frame.}\label{chipc}
\end{figure}

By running this procedure we obtain the eigenvalues given in
Table \ref{pc}. The eigenvalues indicate the variance of PC's
obtained. We can infer from the variances that  the first PC
describes 91\% of the total variance. We assume, therefore, that
the information from the spatial density is concentrated into this
variable. From the first PC we compute the $\chi^2_k$ variables for $df=8$,
$df=10$, $df=12$ and $df=14$ degrees of freedom. The results are
displayed in Figure \ref{chipc}.

\begin{table}
  \centering
  \begin{tabular}{|c|cccc|} \hline
      & PC1 & PC2 & PC3 & PC4 \\ \hline
    eigen val. & {\bf 3.517} & 0.222 & 0.077 & 0.049 \\
    st. dev. & {\bf 1.875} & 0.472 &  0.279 & 0.221 \\ \hline
  \end{tabular}
  \caption{Eigenvalues resulting from the PCA. The values demonstrate
  clearly that only the first PC (marked in bold face) is significant.}
  \label{pc}
\end{table}

\section{Discovery and nature of the GRB ring}
In all frames in Figure \ref{chipc} there is a strong peak
exceeding the $99.9\%$ significance level (99.95\%, 99.93\%,
99.96\%, 99.97\% at $k$=8, 10, 12, 14, respectively) at about 2800
Mpc corresponding to a group of outlying points in Figure
\ref{logd}. This appears to indicate the presence of some real density
enhancement.

\subsection{Discovery of the ring}\label{dring}
\begin{figure}
  \includegraphics[width=8cm]{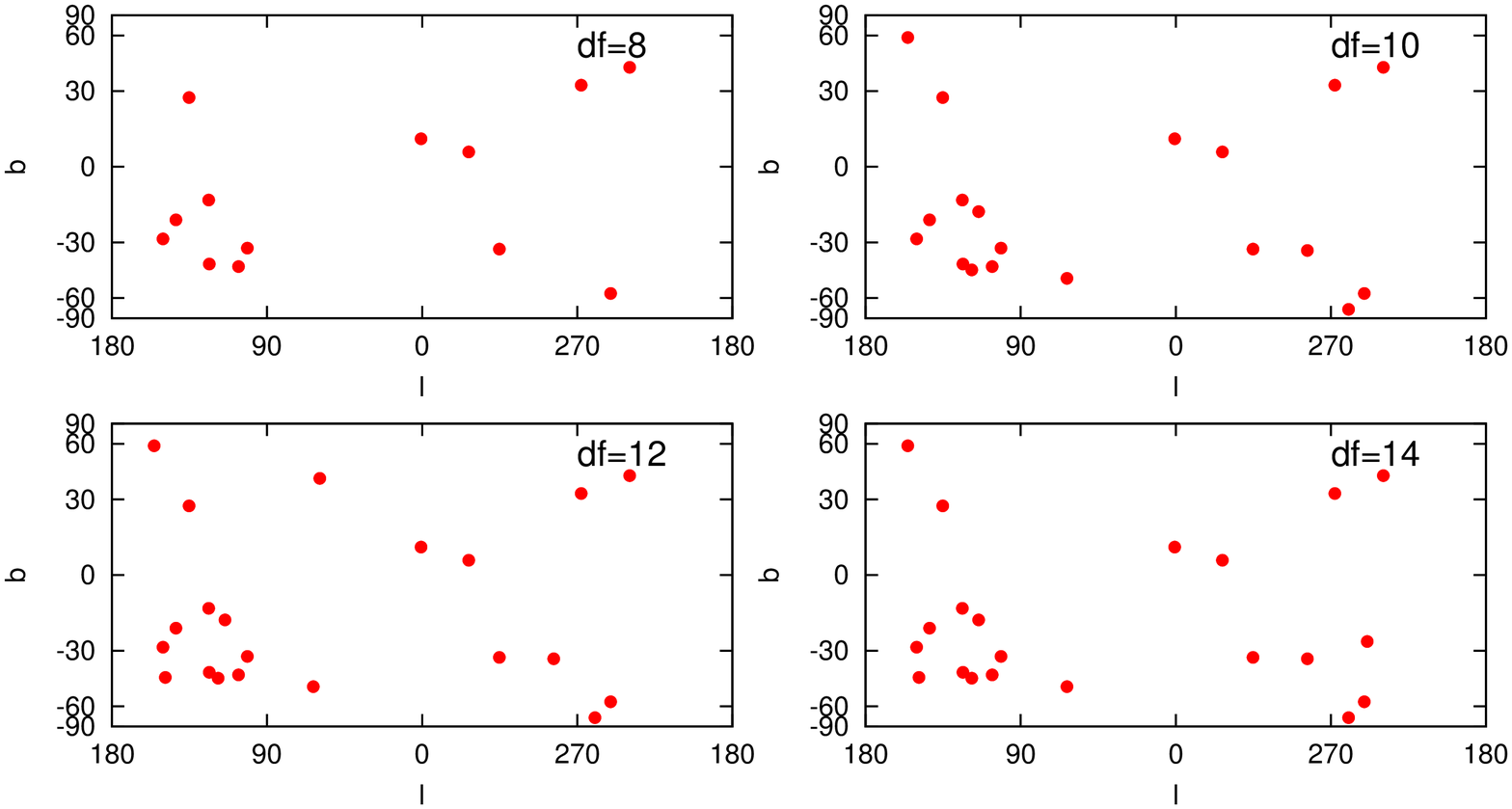}\\
  \caption{Angular distribution of GRBs in the $FWHM$ distance range
  of the highest peaks in Figure \ref{chipc}.  The degree of freedom
  in the upper right corner has the same meaning as in
  Figure \ref{chipc}. Note the ring-like structure of objects in the
  lower left part of the frames.}\label{ring}
\end{figure}

We assume that  GRBs making some contribution  to the highly
significant peak shown in Figure \ref{chipc} lie within the full
width at half maximum (FWHM) angular distance of the peak. Their
angular distribution is shown in Figure \ref{ring}. The most
conspicuous feature in all of the frames is a ring-like structure
in the lower left side of the frames. The redshift, distance and
Galactic coordinates of the GRBs displaying the ring are given in
Table \ref{zdclb}. Using the data listed in the Table one can
calculate the mean redshift and distance of the ring, along with
the standard deviations of these variables, yielding $z=0.822$,
$\sigma_z=0.025$ and $d_c=2770$ Mpc, $\sigma_d=65$ Mpc.

\begin{table}
  \centering\begin{tabular}{|ccccc|}
  \hline
 GRB ID &  redshift & distance (Mpc) & l (deg) & b (deg) \\ \hline
 040924  &  0.859 & 2866 & 149.05 &  -42.52 \\
 101225A  &   0.847 & 2836 & 114.45 & -17.20 \\
 080710  &   0.845 & 2831 & 118.43 & -42.96 \\
 050824  &  0.828 & 2786 &  123.46 & -39.99 \\
 071112C  &  0.823 & 2772 & 150.37 & -28.43 \\
 051022  &   0.809 & 2736 & 106.53 & -41.28 \\
 100816A  &  0.804 & 2723 & 101.39 & -32.53 \\
 120729A  &  0.800 & 2712 & 123.85 & -12.65 \\
 060202  &  0.785 & 2672 & 142.92 & -20.54\\ \hline
  \end{tabular}
  \caption{Redshift, comoving distance, and galactic coordinates of the
  GRBs contributing to the ring-like angular structure. }\label{zdclb}
\end{table}

By definition, the true characteristic physical size $D$ of the
object can be obtained from the $D=\Theta\times d_a = \Theta
\times d_c/(1+z)$ relation, where $\Theta$ is the mean angular
size and $d_a$ is the angular distance. Substituting the
corresponding values obtained above one gets $D=944$ Mpc,
corresponding to 1720 Mpc in the comoving frame.

\subsection{Verification of the ring structure}
In subsection \ref{dring} we claim to find a regular structure in
the shape of a ring. However, the form of this structure is thus
far based only on a visual impression. In the following discussion
we try to give a quantitative value supporting the sensibility of
this subjective impression.

The procedure we used to obtain the very low probability of this
density enhancement only by chance is not sensitive to the true
shape of this clustering. Assuming this clustering to be real we
may compute the probability of getting a ring-like structure only
by chance by assuming some concrete space distribution for the
objects. We make these calculations for the cases of a) a
homogeneous sphere and b) a shell (for more details see the
Appendix).

The probability of observing a ring-like structure is $p=0.2$ in
the case of a shell but it is only $p= 4\times10^{-3}$ for a
homogeneous sphere. It is worth noting that real space
distributions of cluster members generally concentrate
more strongly towards the center than do the elements of
a homogeneous sphere.
Therefore, this probability can
be taken as an upper bound for a probability of obtaining a ring
shape purely by chance.

Combining this latter probability with that of observing the
clustering purely by chance we obtain a value of $p=2
\times\,10^{-6}$ for observing a ring entirely by chance. Thus,
despite the large angular size of the extended GRB cluster, we
find evidence that the cluster represents a large extended ring in
the
 $0.78 \le z < 0.86$ redshift range.

\subsection{The physical nature of the ring}
If we assume that the ring represents a real structure, then we
can speculate about its nature and origins. Perhaps a simple
explanation is that it indicates the presence of a ring-like
cosmic string. This would indicate that it is a large-scale
%One may speculate on the true physical nature of the ring.
%According to a trivial explanation  it is a ring like cosmic
%string also in the reality. In this way it is a large scale
component of the cosmic web, representing the characteristic spatial
distribution of the objects in the Universe.
The main difficulty with this simple
explanation lies with the uniformity of the redshifts (distances)
along the object, indicating that we must be seeing the ring nearly face-on.
This possibility cannot be excluded, but alternative explanations
are also worth of considering.

GRBs are short-lived transient phenomena. The GRBs that
compose the ring, along with their redshifts,
were collected over a period of about ten years. The
number of observed events is determined by the time frequency of
such events in a given host galaxy.
However, the total number of
hosts in the region containing the GRB ring and not having burst events
during the observation period
must be much greater relative to those which were observed.

The number of observed GRBs should be proportional to
the number of progenitors in the same region, although the spatial
stellar mass density is not necessarily proportional to the
spatial number density of progenitors. Namely, the progenitors
for the majority of GRBs are thought to be short-lived $20-40\, M_\odot$
stars, and as such their presence should be strongly dependent on the  star
formation activity in their host galaxies. Thus, our knowledge
of the underlying mass distribution is sensitive to assumptions
about star formation within the ring galaxies.

\subsubsection{Mass of the ring}
\label{mring}

In order to estimate the mass of the ring
structure we make two extreme assumptions providing lower and
upper bounds for the mass. We get a lower bound for the mass
by assuming that the general spatial
stellar mass density is the same in the field and in the ring's
region and only the star formation and consequently the GRB
formation rate is higher here. We get an upper bound for
the mass by supposing a strict proportionality between the stellar
mass density and the number density of the progenitors. For both
estimates we need to know the local stellar density.

Several recent studies have attempted to determine the stellar
(barionic) mass density and its relation to the total Universal
mass density. \cite{Bah2014} determined the stellar mass fraction
and found it to be nearly constant on all scales above $~300 \,
h^{-1}\, kpc$, with $M^*/M_{tot} \cong 0.01 \pm 0.004$.
\cite{LeF2013,LeF2014} issued the VIMOS VLT Deep Survey (VVDS)
final and public data release offering an excellent opportunity to
revisit galaxy evolution. The VIMOS VLT Deep Survey is a
comprehensive survey of the distant universe, covering all epochs
since $z \sim 6$. From this, \cite{Dav2013}  measured the
evolution of the galaxy stellar mass function from $z = 1.3$ to $z
= 0.5$ using the first 53 608 redshifts.

\cite{Mar2013} investigated the dependence of galaxy clustering on
luminosity and stellar mass in the redshift range $0.5 < z < 1.1$
using the ongoing VIMOS Public Extragalactic Survey (VIPERS).
Based on their sample of 10095 galaxies, \cite{Dri2007} estimated
the stellar mass densities at redshift zero amounting to $8.6 \,
h^{-1} \pm 0.6 \times 10^8 M_\odot/Mpc^3$. We use this local value
as our measure of the mass density in the comoving frame and use
it for our subsequent calculations.

We assign a volume to the ring by computing the convex hull (CH)
of the points representing the GRBs in the rest frame. Using the
\textcolor{red}{\tt Qhull}\footnote{{\tt http://www.qhull.org/}
\textcolor{red}{\tt Qhull} implements the {\tt Quickhull}
algorithm for computing the convex hull } program we obtain
a value of $1.9\times 10^8$ Mpc$^3$
for the volume of the CH.
Supposing that the stellar mass density is the same in the ring's
region (i.e. within the CH) as in the field and only the number of
progenitors is enhanced we compute a mass of
$2.3\times10^{17}\, M_{\odot}$ inside the CH.

Alternatively, if we assume that the fraction of progenitors is the
same along the ring as it is in the field, then the total stellar
mass in the volume of a shell with 2770 Mpc radius and 200
Mpc thickness (the observed distance range of the GRBs in the
ring) is $2.2\times 10^{19}\, M_{\odot}$. Since the number of
GRBs making up the ring is about the half the total
observed in the shell, we get a mass of $1.1\times 10^{19}\,
M_{\odot}$ within the ring's CH.

Supposing a strict proportionality between the spatial densities
of the number of GRB progenitors and stellar masses we estimate a
factor of 50 times more mass than that which is obtained above using the field value
within the CH. In the case of a homogeneous mass distribution
within the CH this implies an overdensity by a factor of 50
compared to the field.

In reality, however, the overdensity within the CH appears to be
concentrated in the outer half of its volume in order to produce a
ring-like distribution, suggesting an overdensity enhanced by a
factor of more than 100. This high value appears to be
unrealistic. To resolve this contradiction, the proportion of
progenitors in the stellar mass has to be increased by at least an
order of magnitude and the stellar mass density has to be
decreased by the same factor in the outer half volume of the CH.
This results in a value of $1\times 10^{18} \,M_{\odot}$, which
still represents an overdensity of a factor of 10 suggesting the
ring mass is in the range $10^{17}-10^{18} \,M_{\odot}$, depending
on the fraction of progenitors in the stellar mass distribution.

\subsubsection{The case for a spheroidal structure}
\label{sring}

\begin{figure}
  \includegraphics[width=8cm]{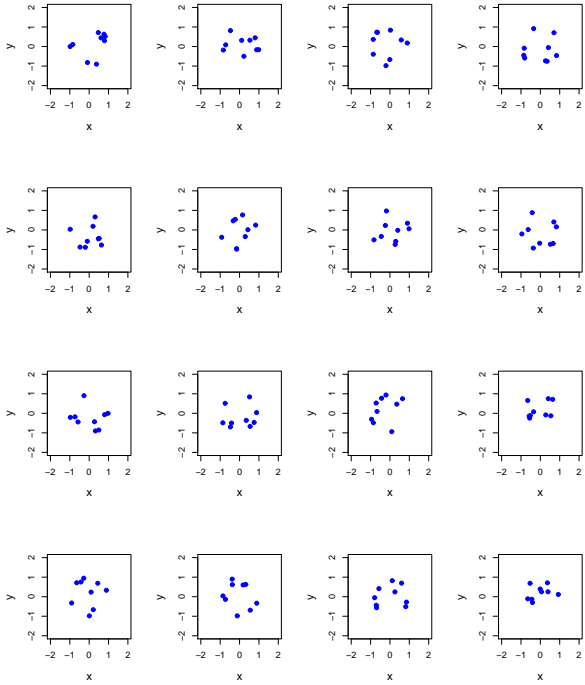}\\
  \caption{Monte Carlo simulation of projecting points into a plain,
  distributed uniformly on a sphere. It is worth noting that some of
  the simulations strongly resemble the observed ring.}\label{MC}
\end{figure}

To overcome the difficulty caused by the low probability of seeing
a ring nearly face on one may assume the GRBs populate the surface
of a spheroid which we see in projection. To demonstrate that the
projection of GRBs uniformly populating the surface of the
spheroid really can produce a ring in projection, we make MC
simulations displayed in Figure \ref{MC}. The simulations show
that a ring structure can be obtained easily by projecting a
spheroidal shell onto a plane. The probability of observing a ring
in this way is much larger than that of observing a ring face on.

Unfortunately, this approach also faces some problems. Assuming
the observed ring is a projection onto a plane, one can calculate
the standard deviation of distances of the objects to the
observer. A simple calculation shows this standard deviation is
about 58\% of  the radius in case of a sphere. Previously we
obtained 1720 Mpc for the diameter of the ring resulting in a 860
Mpc radius with a 499 Mpc standard deviation for the comoving
distances. With the projection correction, however, we obtain only
65 Mpc for this value. This result is obviously in tension with
the value of the standard deviation assuming a spherical
distribution for the GRBs displaying the ring.

The relatively low standard
deviation of the distances, however, is not necessarily caused by
some physical property of the structure but
could be caused by the $FWHM$ of the
statistical signal.  Increasing the distance range around the peak
of the statistical signal relative to the value of the standard deviation
increases the foreground/background as well, and
indicates that the structure may
be buried in the noise. Nevertheless, for the case of a projected
sphere increasing the distance range in this way implies that
the total number of GRBs displaying this structure also has to be
increased by a factor of 6. This would cause the $FWHM$ of the
statistical signal to be much wider, in contrast to what is
observed.

One may resolve this tension somewhat arbitrarily in the following
way. The 499 Mpc value for the standard deviation of the comoving
distances was obtained by assuming a shell-like GRB distribution.
Let us take an interval around the mean distance of 2770 Mpc
within the standard deviation of 499 Mpc. The endpoints correspond
to some lookback time difference between GRBs detected at the same
moment by the observer. The lookback time can be calculated from
the following equation:

\begin{equation}\label{lbt}
t_L(z)=t_H \int^z_0
\frac{dz'}{(1+z')\sqrt{\Omega_M(1+z'^3)+\Omega_\Lambda}}
\end{equation}

\noindent where $t_H=1/H_0$ is the Hubble time. Calculating the
time difference one obtains $\Delta t_L=1.9\times 10^9$ years.
Computing the time difference taking the observed 65 Mpc standard
deviation instead of 499 Mpc, one gets only $\Delta t_L=2.5\times
10^8$ years. If the GRBs displaying the observed ring really
populate a sphere, then the presence of the low 65 Mpc standard
deviation reveals a $2.5\times 10^8$ year period in the life of
the host galaxy when it is very active in producing GRBs.
Furthermore, one has to assume that this happens for all hosts
simultaneously. This coordinated activity may happen by some
external effect which is responsible for the formation of the
sphere.

One can make a similar estimate of the sphere's mass by assuming
that the ring represents a real structure and is not simply a
projection. The difference in this case is that the ring mass
represents only a fraction of the sphere's mass. There are 7
objects in the ring within the 65 Mpc standard deviation. There is
a factor of 7 for getting the number of GRBs within $1\sigma$ on
the sphere. This number represents 68\% of the total number of
GRBs on the sphere, i.e. one should multiply the mass obtained for
the ring by roughly a factor of ten, yielding $10^{18}-10^{19}\,
M\odot$ for the sphere.

\subsubsection{Formation of the ring}
\label{fring}

No matter whether we interpret the spatial structure of the ring
as a torus or as a projection of a spheroidal shell, the formation
of a structure with this large size and mass provides a real challenge
to theoretical interpretations. In addition to the size and the mass
of the structure, one has to explain why the GRB activity is much
higher along the ring than it typically is in the field.

There is general agreement among researchers that following
the early phase of the Big Bang the initial perturbations evolved
into a cosmic web consisting of voids surrounded by string-like
structures. A filamentary structure surrounding sphere-like voids
is a typical result of gravitational collapse
\citep{Cent1983,Ick1984}.

The hierarchy of structures in the density field inside voids is
reflected by a similar hierarchy of structures in the velocity
field \citep{Ara2013}. The void phenomenon is due to the action of
two processes: the synchronisation of density perturbations of
medium and large scales,  and the suppression of galaxy formation
in low-density regions \citep{Ein2011}.

It is generally assumed that the maximal size of these structures
is $100-150$ Mpc \citep{Fri1995,Ein1997,Suh2011,Ara2013}. Quite
recently \cite{Tul2014} discovered the local supercluster (the
"Laniakea") having a diameter of 320 Mpc. This scale is several
times smaller than the estimated 1720 Mpc diameter of the GRB
ring, although perturbations on larger scales cannot be excluded.
However, \cite{Dor1988} have presented arguments that
perturbations on the $200 - 300$ Mpc scale should be excluded.
This value is in a clear contradiction to the existence of the GRB
ring and other large observed structures. Resolution of this
contradiction is still an open issue.

The existence of the ring, either as a torus or the projection of
a spheroidal shell, requires a higher spatial frequency of the
progenitors along the ring than in the field. A possible
interpretation of the higher fraction of progenitors along the
ring is that hosts are still in the formation process
at $6.7\times 10^9$ years after the Big Bang. This supports the
view that large scale structure can form and evolve slowly from
the initial perturbations \citep{Zel1982,Ein2006}.

Dark matter must be given a dominant role in large-scale structure
theories in order to account quantitatively for the formation and
evolution of the cosmic web. That is because the observed
distributions of galaxies are inconsistent with gravitational
clustering theories and with the formation of super clusters in a
wholly gaseous medium \citep{Ein1980}. Recent extensive numerical
studies that include dark matter indicate that its presence
accounts for the basic properties of the cosmic web
\citep{Spr2005,Ang2012}, and these studies have reproduced the
cosmic star formation history and the stellar mass function with
some success \citep{Vog2013}. The very large high-resolution
cosmological N-body simulation, the Millennium-XXL or MXXL
\citep{Ang2012}, which uses 303 billion particles, modeled the
formation of dark matter structures throughout a 4.1 Gpc box in a
$\Lambda CDM$ cold dark matter cosmology.  \cite{Kim2011}
presented two large cosmological N-body simulations, called
Horizon Run 2 (HR2) and Horizon Run 3 (HR3), made using $6000^3 =
216$ billions and $7210^3 = 374$ billion particles, spanning a
volume of $(7.200 \, h^{-1}Gpc)^3$ and $(10.815 \, h^{-1}Gpc)^3$,
respectively. Although these  sizes of the simulated volumes were
large enough to produce very large structures, accounting for
local enhancements corresponding to the size of the GRB ring
structure is still an open problem.  We address this issue in the
next subsection.

\subsubsection{Spatial distribution of GRBs and large scale
structure of the Universe}

 In subsection \ref{fring} we noted that very large scale
cosmological simulations may account for huge disturbances in the
dark matter distribution, in particular for   LQGs and the object
discovered by \cite{Hoi2013}. Nevertheless, in subsections
\ref{mring} and \ref{sring} we pointed out that the existence of
the GRB ring probably can not be accounted for a simple
enhancement of the underlying barionic and dark matter density.
Presumably, to explain the existence of the ring one needs a
coordinated enhanced GRB activity in the responsible host
galaxies.

According to a widely accepted view the majority of the observed
GRBs are resulted in collapsing high mass ($20-40 \, M_\odot$)
stars. GRBs are very rare transient phenomena, consequently, they
observed spatial distribution is a serious under sampling of the
space distribution of galaxies in general. Furthermore, the high
mass stars have  short lifetimes, consequently GRBs prefer
those galaxy hosts having considerable star forming activity.

Due to their immense intrinsic brightnesses, GRBs can be detected at
large cosmological distances.  GRB 090423 has $z=8.2$, the largest
spectroscopically measured redshift \citep{Tan2009}. Even though
GRBs seriously under-sample the matter distribution, they are the
only observed objects doing so for the Universe as a whole up to
the distance corresponding to the largest measured redshift.

Since there is no complete observational information on the
spatial distribution of dark and barionic matter on the same scale
as that of GRBs, one has to use the large scale simulations of
the distribution of the cosmic matter for making such comparisons.
We used for this purpose the publicly available Millennium-XXL
simulation\footnote{\tt http://galformod.mpa-garching.mpg.de/portal/mxxl.html}.

As we mentioned at the end of subsection \ref{fring}, the 4.1
Gpc size of the simulated volume is large enough to account for
structures with characteristic size of the ring. Since GRBs prefer
host galaxies with high star formation activity, we calibrate the
dark matter density in XXL to the spatial number density of
galaxies having large star forming rate ($SFR$) assuming that

\begin{equation}\label{dmsfr}
    \nu_{s}(x,y,z)) = c(\varrho_{d})\varrho_{d}(x,y,z)
\end{equation}

\noindent where $\nu_{s}$  represents the spatial number density
of star forming galaxies and $\varrho_{d}$ the density of the dark
matter. We assume that the $c(\varrho_{d})$ conversion factor
depends only on $\varrho_{d}$ but not on the spatial coordinates
and that it is identical in the XXL and the Millennium simulation.

We determine the $c(\varrho_{d})$ conversion factor using the
data available in the Millennium simulation. The GRB ring is
located in the $0.78 < z < 0.86$ redshift range, so we select
the $z=0.82$ slice of the simulation.  The star forming galaxies
have $SFR > 30\,  M_\odot \, yr^{-1}$ at this redshift
\citep{Per2015}.

\begin{figure}
  \includegraphics[width=5cm,angle=270]{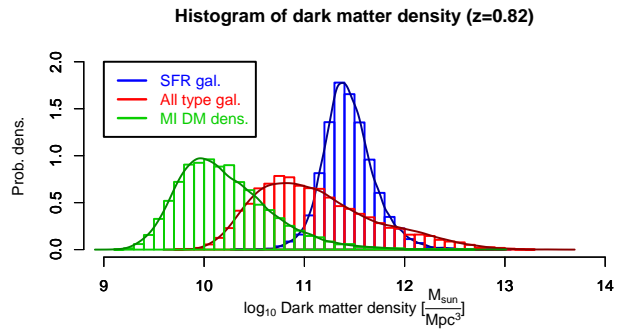}\\
  \caption{Probability distribution of the dark matter density
  in the Millennium simulation. Distribution of dark matter density in the
  simulation (green), for galaxies in general (red), and
  for star forming galaxies (blue). The star forming galaxies prefer
  a certain range of underlying dark matter density that differs
  from that of galaxies in general.}\label{dmdens}
\end{figure}

As one can infer from Figure \ref{dmdens}, the selected star
forming galaxies prefer a certain dark matter density range:
for densities less than and greater this range such galaxies
are uncommon in the sample.  This
range differs from that of galaxies in general.  This may
indicate that the spatial distribution displayed by the galaxies
in general is not necessary identical with that shown by the
GRBs.

After determining the $c(\varrho_d$) conversion factor, we obtain
from Equation (\ref{dmdens}) the $\nu_s$ spatial number density of
the star forming galaxies in the XXL simulation. Based on this
spatial distribution we generate random samples of sizes
comparable to that of the observed GRB frequency. For generating the
simulated sample we use the Markov Chain Monte Carlo (MCMC)
method implemented in the \textcolor{red}{\tt metrop()} procedure
available in {\bf R}'s {\it mcmc} library.

An important issue for using this algorithm is to check whether or not the
simulated Markov chain has reached its stationary stage. We
check it by computing the auto regression function of the
simulated sample  by the \textcolor{red}{\tt acf()} function in
{\bf R}. We also check the MCMC output by conventional MC.

The known number of GRBs now exceeds a couple of thousand and is
steadily increasing with ongoing observations. Unfortunately, only
a fraction of these have measured redshifts. Motivated by the
number of known GRBs and by their relationship to star forming
galaxies, we make MCMC simulations of the $\nu_{s}(x,y,z)$ spatial
number density of these galaxies from Equation (\ref{dmdens}),
getting sample sizes of 1000, 5000, 10000 and 20000.

\begin{figure}
  \includegraphics[width=6cm,angle=270]{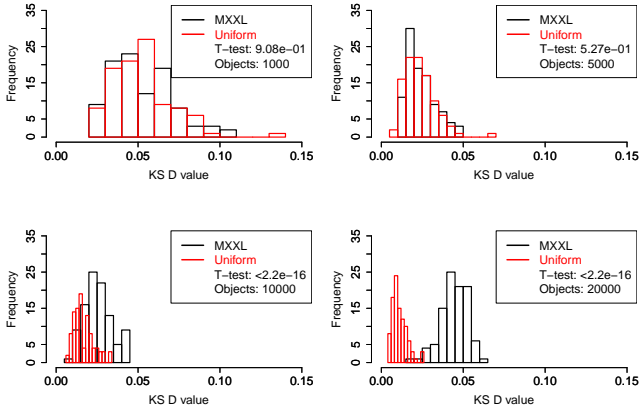}\\
  \caption{Comparison of the KS differences between the XXL  and CSR
  (black) and the CSR (red) cumulative
  distributions of the $k=12^{th}$ nearest neighbors distances.
  Note the significant differences between the XXL and CSR case at
  sample sizes of N=10000 and 20000 (bottom left and right) unlike
  to N=1000 and 5000 (top left and right) where there are no
  significant differences.
  }\label{KSD}
\end{figure}

We make 100 simulations for these sample sizes and compare
them with completely spatially random (CSR) samples of the same
sizes in the XXL volume. In subsection \ref{knear} we computed the
nearest neighbors of the $k=8,10,12$ and $14$ order. Following the
same procedure here, we obtain the nearest neighbor distances of the
$k=12^{th}$ order for the XXL and the CSR samples and using the
\textcolor{red}{\tt ks.test()} procedure in {\bf R}'s {\it stats}
library, and compute the maximal difference between the cumulative
distributions. We repeat this procedure between the CSR nearest
neighbor distributions. The distributions of KS differences
between XXL-CSR and CSR-CSR samples are displayed in Figure
\ref{KSD}.

As one may  infer from this figure, the distribution of KS
differences between XXL-CSR samples do not differ significantly
from those of CSR-CSR in the case of the sample sizes of $N=1000$
and 5000. On the contrary, in the case of $N=10000$ and 20000 the
difference between the XXL-CSR and CSR-CSR cases is very
significant.

Based on this result one may conclude that the simulated XXL
samples with sizes of $N=1000$ and 5000 do not differ from the CSR
case. On the other hand, the samples with sizes of $N=10000$ and
20000 differ significantly from the CSR case. Each  sample corresponds to some
mean distance to the nearest object of the $k=12^{th}$ order. The
computed mean distances are tabulated in Table \ref{pcsr}.

Obviously, groups having a characteristic size of 280 Mpc can
be detected with a sufficiently large sample size. This value
corresponds to the largest structure (251 Mpc) found by
\cite{Par2012} using the HR2 simulation.  The number of GRBs
detected, however, is insufficient for revealing this scale.
Consequently, if the XXL simulation correctly represents the
large scale structure of the Universe the GRBs reveal it as
CSR on the scale corresponding to the sample size.

At this point, however, it may be appropriate to repeat the remark
made at the beginning of this subsection: the existence of the GRB
ring can not be explained by a simple density enhancement of the
underlying barionic and/or dark matter. It probably needs some
coordinated star forming activity among the responsible GRB hosts.
In this case the spatial  distribution of GRBs does not
necessarily trace the underlying matter distribution in general.
However, we can not exclude the possibility that the XXL
simulation does not correctly account for all possible large scale
structures and GRBs are mapping a structure that was not
simulated.

\begin{table}
  \centering
  \begin{tabular}{ccc}
    \hline
    Sample size & Mean dist. [Mpc] & Prob. of CSR \\
    \hline
    \ 1000 & 627 & 0.39 \\
    \ 5000 & 351 & 0.71 \\
    {\bf 10000} & {\bf 277} & ${\bf <2.2e-16}$ \\
    {\bf 20000} & {\bf 217} & ${\bf <2.2e-16}$ \\
    \hline
  \end{tabular}
  \caption{Mean distances to the $k=12^{th}$ order nearest
  neighbor of star forming galaxies at different sample sizes
  in the XXL simulation and the   probability for being CSR.
  The significant deviation from the   CSR case is marked in
  bold face. Note that this size is consistent  with the CP
  and more than six times smaller than  the GRB ring
  in this paper.}\label{pcsr}
\end{table}

Some concern may arise, however, concerning the interpretation of
the ring as a true physical structure, and of the causal
relationship between the GRBs displaying it. \cite{Suh2011} has
pointed out that cosmic structures greater than 140 Mpc in
co-moving coordinates did not communicate with one another during
the late stage of universal expansion preceding Recombination. The
skeleton of the web was created during the inflationary period
\citep{kof1988} and evolved slowly following this epoch.

The volume of the shell between $0.78 < z < 0.86$ is
$20.2 \,Gpc^3$ in the co-moving frame. The corresponding
volume is $2.1 \,Gpc^3$ for $z = 0.2$ in
the case of the SDSS main sample
and $14.4 \,Gpc^3$ for $z = 0.4$ for luminous red galaxies (LRGs),
respectively. The volume of the shell is about ten thousand times
larger than the volume of a typical supercluster found by
\cite{Lii2012} in the SDSS data.

Since the cosmic web evolves slowly, the structure of the GRB ring
should exhibit the same general characteristics as those displayed by
superclusters defining the web.
Comparing the estimated number of superclusters with
the number of detected GRBs, it appears that every thousand
superclusters has produced on average one measured GRB.
GRBs are therefore very rare events superimposed on the web,
and the small probability of GRB
detection casts serious doubt on the nature of the
GRB ring as a real physical structure.

Taking these distributional characteristics into account suggests
that the Ring is probably not a real physical structure. Further
studies will be needed to reveal whether or not the Ring structure
could result from a low-frequency spatial harmonic of the
large-scale matter density distribution and/or of universal star
forming activity.

\section{Summary and conclusion}
Motivated by the recent discovery of \cite{Hoi2013} revealing a
large Universal structure displayed by GRBs, we study the
spatial distribution of these objects in the comoving frame. The
advantage of this approach is that GRBs (which are
short transients) have footprints in this frame that do not change
with time.

We assume in this approach that, for the spatially homogenous and
isotropic case, the joint observed distribution of the GRBs  can be
factored into two parts: one part depends on the angular
coordinates while the other part is radial and depends on the redshift.

This assumption can be tested in two different ways. The first
method is essentially that used by
\citetalias{Hoi2013} which compares the conditional probability of
the GRB angular distribution at different $z$ values.
The second method tests whether resampling the GRBs randomly
makes any statistical changes in the distribution in the 3D
comoving frame.

We estimate the spatial density of GRBs by searching the
angular separations
of the $k$-th order nearest neighbours. For these
computations we use the \textcolor{red}{\tt knn.dist(x,k)}
procedure in the {\it FNN} library of the {\bf R} statistical
package. To compromise between the large variance of
estimated densities at small $k$ values and the smearing out of real small
scale structures at large $k$ values, we use the spatial densities
obtained by taking $k=8,10,12$ and $14$.

Resampling the redshift distribution 10000 times and calculating
the spatial densities from the samples obtained in this way we
obtain mean densities and their variances assuming the null
hypothesis, i.e. that the factorization of the spatial distribution of
GRBs is valid. Subtracting the mean value from the observed one
and dividing by the standard deviation we calculate the
standardized values of the densities obtained from the nearest
neighbour procedure.

KS tests revealed that the logarithmic densities obtained in this
way follow a Gaussian distribution allowing us to get a
logarithmic likelihood function as a sum of the squared
logarithmic densities. Since the sum of squared Gaussian variables
follows a $\chi^2_k$ distribution with $k$ degrees of freedom, by
selecting objects in a certain distance separation range and
calculating the value of this variable we can test for the
significance of density fluctuations.

Since the calculated logarithmic densities in the $k=8,10,12,14$
cases are strongly correlated pairwise, performing a principal
component analysis (PCA) allows us to join the logarithmic
densities in the first (the only significant) PC variable
representing 91\% of the total variance. Computing  $\chi^2_k$
values from this PC for $k=8,10,12,14$ degrees of freedom and
plotting them  as a function of the distance, we find a very
pronounced peak at about 2800 $Mpc$ corresponding to a
significance of 99.95\%, 99.93\%, 99.96\% and 99.97\%, depending
on the degrees of freedom.

We plot the angular positions of the GRBs within the $FWHM$
range around the distance of the $\chi^2_k$ peak. Examining
these plots we conclude the following:

 \begin{itemize}
    \item[-] There is a ring consisting of 9 GRBs having a mean
    angular diameter of $36^o$ corresponding to 1720 Mpc in the comoving frame.
    \item[-] The ring is located in the $0.78<z<0.86$ redshift range
    having a standard deviation of  $\sigma_z=0.025$,
    corresponding to a comoving distance range of $2672<d_c<2866$ Mpc
    having a standard deviation of $\sigma_d=65\, Mpc$.
    \item[-] If one interprets the ring as a real spatial structure, then
    the observer has to see it nearly face on because of the small
    standard deviation of GRB distances around the object's center.
    \item[-] The ring can be a projection of a spheroidal structure.
    Adopting this approach one has to assume that each host galaxy has
    a period of $2.5 \times 10^8$ years during which the
   GRB rate is enhanced.
    \item[-] The mass of the object responsible for the
    observed ring is estimated to be in the range of
    $10^{17}-10^{18}\,M\odot$ if the true
    structure is a torus or $10^{18}-10^{19}\,M\odot$
    in case of a  spheroid.
    \item[-]  GRBs are very rare events superimposed on the cosmic web
    identified by superclusters.  Because of this, the ring is probably not a real
    physical structure. Further studies are needed to reveal
    whether or not the Ring could have been produced by a low-frequency spatial
    harmonic of the large-scale matter density distribution
    and/or of universal star forming activity.

 \end{itemize}

 \section*{Acknowledgements}
 This work was supported by the OTKA grant NN 11106 and NASA
 EPSCoR grant NNX13AD28A.  We are very much indebted to the referee, Dr. Jaan
 Einasto, for his valuable comments and suggestions.

\bibliography{Balazs_GRB_ring}

\appendix
\section{Observing a ring structure by chance}
In this manuscript we have found strong evidence for a ring-like
structure displayed by 9 GRBs. The probability of obtaining this
clustering only by chance is about $p=5\times10^{-4}$, but this
value is not sensitive to the actual pattern of the points within
the group. Although we claim to have found evidence for a regular
structure, the apparent shape of a ring is based only on a visual
impression. It is useful to develop a quantitative measure
supporting the efficacy of this subjective impression.

In this appendix we compare the projection of two simple spherical
models; a) a homogeneous sphere and b) a spherical shell. It is
not difficult to derive the probability density functions for
these projections into a plain. If we denote the projected
distance from the mean position of the group to one of the members
by $\varrho$, then we can normalize each position relative to the
maximum projection $\varrho_{max}$ so that the projections vary in
the range of \{0,1\}. In case of a homogeneous sphere the
projected probability density is given by

\begin{equation}
f(\varrho)=3 \varrho \sqrt{1-\varrho^2}
\end{equation}

\noindent and in the case of a shell we get

\begin{equation}
g(\varrho)=\frac{\varrho} {\sqrt{1-\varrho^2}}
\end{equation}

\begin{figure}
  \includegraphics[width=8cm]{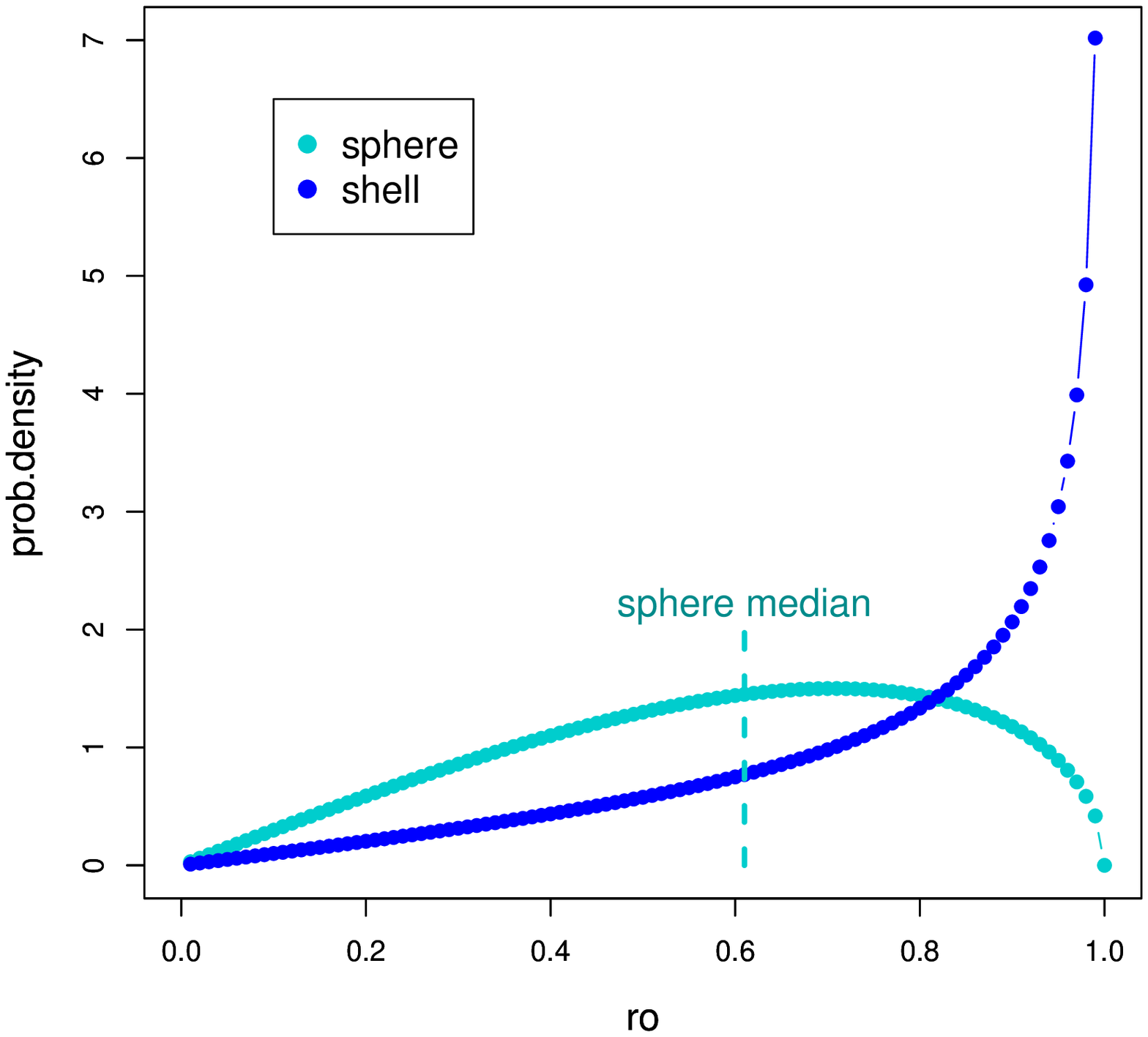}\\
  \caption{Comparison of the projected probability densities of a
homogeneous sphere and a shell. The median of the sphere is
indicated.}\label{sphr}
\end{figure}

\noindent The shapes of these functions are displayed in Figure
\ref{sphr}.

The Figure demonstrates that the projections for a shell
result in a
significant enhancement of the points close to the maximal
distance from the center; this is not the case for a
projected homogenous sphere. We can compute the
probability of measuring all 9 points outside of the median distance. By
definition the median splits the distribution into two parts of
equal probability. The value of the median distance for a projected
homogeneous sphere is $\varrho_{median}= 0.61$.

We can calculate the probability of measuring all 9 points in
the ${0.61 < \varrho < 1}$ regime. To calculate the probability of
this case we invoke the \textcolor{red}{\tt binom.test()}
procedure available in the {\bf R} statistical package. The
probability of finding an object outside the median
distance is $p = 0.5$ by definition. The probability for finding all 9 points
outside the median is given by the binomial distribution.

Similarly we can calculate the probability of having all 9
objects in a region outside the $\varrho_{median} = 0.61$ median,
assuming that the true spatial distribution of the points is a
shell. Integrating the $g(\varrho)$ probability density in the
\{0.61, 1\} interval we get

\begin{equation}\label{med}
    \int\limits^1_{0.61}g(\varrho)=\int\limits^1_{0.61}\frac{\varrho}{\sqrt{1-\varrho^2}}=0.7924
\end{equation}

\noindent Inserting this probability into the binomial test
expression we get $p=0.2192$.

Summarizing the results of the tests performed above, we
infer that the probability of observing a ring-like structure from a
projected 3D homogeneous density enhancement only by chance is
 $p=3.9\times10^{-3}$ while assuming a projected 3D shell it
is much higher, $p=0.22$. This gives us a good reason to believe
that the ring is a result of a projected 3D shell-like regular
pattern.

\end{document}